\def\a{\alpha}
\def\d{\delta}

\def\r{\rho}

\def\ra{\rightarrow}

\documentclass[12pt]{article}

\usepackage{a4wide}

\usepackage{amsfonts}

\begin{document}
\title{Cut-out sets, fractal voids and cosmic structure}
\author{Jos\'e Gaite\\                
{\small Instituto de Matem{\'a}ticas y F{\'\i}sica Fundamental,
CSIC, Serrano 113bis, 28006 Madrid, Spain}
}
\date{March 21, 2006}

\maketitle

\begin{abstract}
``Cut-out sets" are fractals that can be obtained by removing a
sequence of disjoint regions from an initial region of $d$-dimensional
euclidean space. Conversely, a description of some fractals in terms
of their void complementary set is possible.  The essential property
of a sequence of fractal voids is that their sizes decrease as a power
law, that is, they follow Zipf's law.  We prove the relation between
the box dimension of the fractal set (in $d \leq 3$) and the exponent
of the Zipf law for {\em convex} voids; namely, if the Zipf law
exponent $e$ is such that $1 < e < d/(d-1)$ and, in addition, we
forbid the appearance of {\em degenerate} void shapes, we prove that
the corresponding cut-out set has box dimension $d/e$ ($d-1 < d/e <
d$).  We explore the application of this result to the large scale
distribution of matter in cosmology, in connection with ``cosmic
foam'' models.
\bigskip

%\PACS{
\small PACS:
{05.45.Df} {Fractals};  %\and
{02.50.-r} {Probability theory, stochastic processes, and statistics};
%\and
{98.65.Dx} {Galaxy groups, clusters, and superclusters; large
scale structure of the Universe}     
%} % end of PACS codes
\end{abstract}

\section{Introduction}
\label{intro}

Some fractals can be obtained by removing an infinite sequence of
disjoint regions from an initial set.  Mandelbrot \cite{Mandel}
introduced the concept of fractal holes (under the greek word {\em
tremas}) and showed that the distribution of one-dimensional holes
(gaps) follows a simple power law; namely, the number of gaps of
length $U$ greater than $u$ is $N(U > u) \propto u^{-D}$, where $D$ is
the fractal dimension.  Falconer \cite{Falc} has studied the fractal
properties of ``cut-out sets" in terms of box dimensions. In
particular, he has proved that fractal sets in one dimension, that is,
fractal subsets of $\mathbb{R}$, have box dimensions that depend on
the size of the complementary intervals and not on their
arrangement. To be precise, the fractal $E$, with a sequence of void
intervals $a_k$ ($k=1,2, \ldots$) has
$$
\dim_B E = -1/\lim_{k \ra\infty} (\log a_k/\log k) 
$$
if and only if the limit exists.  Roughly speaking, this theorem says
that fractality in $\mathbb{R}$ is related with Zipf's power law
\cite{Zipf} for the rank-ordering of void intervals (with exponent
$1/\dim_B E$).  In turn, Zipf's law is equivalent to the law $N(U > u)
\propto u^{-D}$ if we identify $D$ with $\dim_B E$.

Mandelbrot and Falconer have also considered the generalization to
higher dimension, that is, to fractal subsets of
$\mathbb{R}^d$. However, a void in a subset of $\mathbb{R}^d$ is not
only defined by its size, so the problem is more complex. One can
prescribe a definite shape of voids, for example, discs in $d=2$, so
that the size determines the void (up to its position). Then it still
holds that the box dimension of a cut-out set depends only on the
sequence of sizes of the complementary intervals and does not depend
on their arrangement. However, whereas void regions can be freely
rearranged in $d=1$, in $d > 1$ there are constraints. Examples of
two-dimensional cut-out sets are the Apollonian packings of disks
\cite{Mandel,Falc}.

In addition, while in $d=1$ there is no restriction on the fractal
(Hausdorff-Besicovitch or box) dimension of cut-out sets, except that
it be between zero and one, in $d > 1$, cut-out sets must have {\em
topological codimension} one, so they are formed by curves in two
dimensions or surfaces in three dimensions (etc.) which enclose the
voids.  Therefore, their fractal dimension must be larger than $d-1$.

On the other hand, it is possible to define a sequence of void regions
in a finite approximations to a fractal set by algorithmic methods
that exploit the recursive structure of fractals
\cite{Gaite,Gaite2}. These algorithms are called {\em
void-finders}. The question is to determine the relation between the
voids thus found and the actual fractal set; for example, to discern
in what sense (if any) a general fractal can be considered a cut-out
set of the voids thus found.

The motivation to construct void-finders has arisen in cosmology, but
the interest of fractals for cosmology is older.  It rests on the fact
that the large scale distribution of matter is arguably scale
invariant (within some limits).  The fractal properties of the
distribution of galaxies have been studied for many years (recent
references are in \cite{Piet-Marti}).  The existence of large voids in
the distribution of galaxies, as a counterpart of galaxy clustering,
is an observational fact. The distribution and properties of those
voids have become a subject of systematic study
\cite{cosmo}.  The understanding of voids as a fractal feature of the 
distribution of galaxies motivated us to study general properties of
fractal voids \cite{Gaite,Gaite2}.  In particular, the already
compiled catalogues of voids in the distribution of galaxies are
suitable for standard rank-ordering techniques \cite{Zipf,Sornette} in
which voids are ranked by size.  In Ref.\ \cite{Gaite} we defined
voids as having constant shape: discs or squares, in two dimensions
(the disc case is connected with Apollonian packings).  This
definition is adequate for mathematical treatment, but we found that
it is not satisfactory for analysing a general fractal: often
equal-shape voids of similar size touch each other and visually it
seems that they should be merged into a unique void of {\em irregular}
shape.  This suggested us to improve the definition of void by
considering voids of arbitrary shape. In particular, in Ref.\
\cite{Gaite2}, we devised a void-finding algorithm based on discrete
geometry methods, namely, Delaunay and Voronoi tessellations. It
produces a sequence of voids of polygonal but otherwise arbitrary
shape (in $d=2$).  The rank-ordering of the voids of suitable examples
of fractal point sets in one and two dimensions showed that Zipf's
power-law holds; but we did not attempt to provide any mathematical
proof of this.

So the purpose of this work is: (i) to extend the notion of cut-out
set in $d > 1$ to void regions of non-constant shape and to establish
the relation of the box dimension of the set with the Zipf law for
voids, following Falconer's methods \cite{Falc}; (ii) to see in
what sense and to what extent a general fractal can be considered a
cut-out set of the voids found by algorithmic methods; (iii) finally,
to indicate how to apply our conclusions to the large scale
distribution of matter in cosmology.

To achieve the first goal (i), we will need to introduce concepts of
integral geometry \cite{geom}. These concepts, in particular, the
Minkowski functionals, have already been applied to fractal
distributions by Mecke \cite{Mecke}. His point of view, based on
clusters, can be considered dual to the point of view that we adopt
here, based on voids.

\section{Cut-out sets in more than one dimension}
\label{sec:1}

Cut-out sets are obtained by removing an infinite sequence of disjoint
regions from an initial set.  In one dimension, a cut-out set is a
subset of $\mathbb{R}$ obtained by removing an infinite sequence of
disjoint open intervals from an initial closed interval such that the
sum of the lengths of the removed intervals converges to the total
length of the initial interval.  A trivial example is the Cantor set.
Let the fractal $E \in \mathbb{R}$ be the result of removing from the
closed interval $A$ a sequence of open disjoint intervals
$\{A_k\}_{k=1}^\infty$ such that the length of $A_k$ is $a_k$ and
$\sum_{k=1}^\infty a_k$ equals the length of $A$.  It is not difficult
to see that the set obtained by removing a finite number of intervals
of decreasing length is related with a neighbourhood of $E$ and so the
pattern of sizes of removed intervals is related with the behaviour of
the $r$-neighbourhood of $E$ as $r \ra 0$. If this behaviour is a
power law, it defines the Minkowski-Bouligand dimension, which is
equivalent to the box dimension \cite{Mandel,Falc2}.  A careful
analysis then shows that the corresponding condition on the lengths of
removed intervals is that they follow a power law of their rank, that
is, that they follow Zipf's law \cite{Zipf}.  To be precise, the
fractal $E = A - \cup_{k=1}^\infty A_k$, with the decreasing sequence
of lengths $\{a_k\}_{k=1}^\infty\,,$ has box dimension
$$
\dim_B E = -1/\lim_{k \ra\infty} (\log a_k/\log k) 
$$
if and only if the limit exists. 

In more than one dimension, a cut-out set is a subset of
$\mathbb{R}^d$ obtained by removing an infinite sequence of disjoint
connected open regions $\{A_k\}_{k=1}^\infty$ from an initial compact
region $A$, which is natural to choose convex \cite{Falc}. We need to
restrict the possible shapes of the $A_k$.  The simplest option is to
prescribe a definite shape for them.  Before we analyse the problem in
detail, let us consider a particular type of fractals with voids of
constant shape.

\subsection{Cantor-like fractals and merging criteria}
\label{Cantor}

For Cantor-like fractals, which are strictly self similar, 
Zipf's law for voids is a consequence of their construction 
\cite{Gaite}. A (deterministic) Cantor set is constructed with 
a generator characterized by two numbers, namely, the scaling factor
$r$ and the number of pieces $N$ to remain, and by the arrangement of
these remaining pieces. In one dimension, we convene that if the
arrangement of the removed $1/r - N$ {\em open} intervals is such that
two or more are adjacent then one also removes the isolated points
between them ({\em merging} criterium).  This criterium leads to
characterize the fractal generator by one more number: its number of
voids (gaps) $m \leq 1/r - N$ \cite{Gaite}.  If we do not apply the
merging criterium, the resulting countable set of isolated points does
not contribute to the Hausdorff-Besicovitch dimension of the
Cantor-like set but may contribute to its box dimension \cite{Falc2}.

In $d$ dimensions, the merging criterium generalizes to the removal of
the $d-1$-dimensional boundaries between adjacent open
$d$-cubes. However, a complication arises: it is also possible that
some open $d$-cubes touch the boundary of the initial closed
$d$-cube. If we also convene to remove the corresponding
$d-1$-dimensional boundaries, then, when we iterate the generator, we
merge voids of different levels.  Actually, all the void regions merge
and form one connected void region.  We may not allow this by
preserving in the generator the boundary of the initial closed
$d$-cube. This boundary has topological codimension one, so the box
and Hausdorff-Besicovitch dimensions of the resulting Cantor-like set
are larger than $d-1$.  This holds even when the expected value of the
box and Hausdorff-Besicovitch dimensions, $-\log N/\log r$, is smaller
than $d-1$, so then the boundary preserving criterium alters them
significantly. Even if the expected Hausdorff-Besicovitch dimension is
larger than $d-1$, so that it may not be altered by this criterium,
the structure of the resulting Cantor-like set may change
significantly. 

For example, consider the two-dimensional middle third
Cantor set, constructed as the cartesian product of two middle third
Cantor sets and with dimension $2 \log 2/\log 3 > 1$; it has zero
topological dimension, that is, it is a set of points rather than
lines. Its construction with the two-dimensional generator consisting
of removing the appropriate pattern of five open squares from the unit
closed square with the boundary preserving criterium, makes it a set
of lines (with the same box and Hausdorff-Besicovitch dimensions as
the point set).  This happens irrespective of any criterium for
merging accross inter-cube boundaries: if no merging is prescribed,
then there are additional lines.

Of course, if the generator does not include any open $d$-cube that
touches the boundary of the initial closed $d$-cube, cubes removed in
different iterations cannot merge.  This condition ensures that the
fractal set has topological codimension one.  Typical fractals
generated in this form are the Sierpinski carpets \cite{Mandel}.
Assuming inter-cube merging, it is convenient to introduce the number
$m$ of voids of the generator, such that $1 \leq m \leq 1/r^d - N$.
The Hausdorff-Besicovitch dimension $D=-\log N/\log r > d-1$ is
independent of $m$ and the sizes $a_k$ follow a power law with
exponent $-d/D$ (on average) \cite{Gaite,Gaite2}.

Note that the inter-cube merging criterium implies that voids in the
generator may not have a common shape but a finite number of shapes
instead.  However, this finite number of void shapes is conserved in
the iterations, which only change their size.

\subsection{Voids in general fractals and cut-out sets}

We have seen in the preceding examples of Cantor-like fractals that a
precise definition of voids demands them to have a precise boundary
and, therefore, the corresponding fractal set has topological
codimension one, since it contains the boundaries which have
themselves codimension one. Fractals of topological codimension one
are curves in $d=2$, surfaces in $d=3$, etc. One may wonder if a
codimension one set has an associated set of voids that make it
equivalent to a cut-out set, and if those voids fulfill Zipf's
law. Since we take the initial compact region $A$ of a cut-out set to
be convex, we may consider the convex hull of the fractal and the
complementary set of the fractal with respect to it. If it is formed
by an infinite set of connected regions, this is the natural set of
voids. For example, this approach works for the von Koch curve, whose
set of voids can be shown to fulfill Zipf's law by relying on its
self-similarity.  However, these voids have themselves fractal
boundary and, therefore, are not sufficiently simple for our purposes.

A different approach is algorithmic, namely, to define fractal voids
as the regions found by void-finders. These regions may have constant
shape, being the obvious shapes a square or a disc \cite{Gaite}. Or we
may allow arbitrary shapes; in particular, we have devised a new
void-finder, based on discrete geometry methods (Delaunay and Voronoi
tessellations), that produces a sequence of voids of arbitrary
polygonal or polyhedral shape, respectively, in $d=2$ or in $d=3$
\cite{Gaite2}.

The problem with the constant shape voids found in an arbitrary
fractal is that they may not fill its real voids of variable shape, if
they exist. For example, let us assume that a fractal in $d=2$ has a
square void that we are trying to fill with discs. This filling will
form the Apollonian packing of the square, which leaves out a fractal
region of dimension about 1.31 \cite{Falc}.  Then it may happen that
this dimension is larger than the dimension of the original fractal
with the square void. In contrast, arbitrary polygonal (or polyhedral)
shapes are adaptable. Therefore, if a fractal has topological
codimension one and so it has well-defined voids, appropriate
polygonal (or polyhedral) shapes will reproduce them with sufficient
approximation. However, this approach still allows for too general
void shapes because the voids may have fractal boundary and diverging
perimeter length (in $d=2$, or diverging surface area in $d=3$).

Arbitrary polygonal (or polyhedral) void shapes seem a suitable
starting point but they are still too general. We would like them to
have a bounded perimeter length (or surface area) for a given {\em
diameter} (we define diameter as the greatest distance apart of pairs
of points \cite{Falc2}). A nice way of implementing this condition
without being too restrictive, that is, of preventing the boundary of
voids from becoming wrinkled, is to require them to be convex.  So we
define henceforth a cut-out set as a subset of $\mathbb{R}^d$ obtained
by removing an infinite sequence of disjoint open convex
$d$-polyhedral regions $\{A_k\}_{k=1}^\infty$ from an initial compact
convex region $A$.  In the end, the restriction to $d$-polyhedra
proves to be superfluous and we may consider general convex
voids. However, from a computational point of view, that is, when
finding voids in a finite sets of points, these voids must be
polyhedra.  In the next section, we generalize the results of Falconer
regarding the box dimension of fractals resulting from cutting out
discs \cite{Falc} to these more general cut-out sets, restricting
ourselves to $d=2$ as well. Further generalization to $d=3$ is
achieved in the following section.

\subsection{Two-dimensional cut-out sets with convex polygonal voids}

Let $A$ be a plane compact convex region of perimeter $p$ and let
$\{A_k\}_{k=1}^\infty$ be a sequence of disjoint open convex polygonal
regions, such that $A_k$ has diameter $\d_k$, perimeter $p_k$, and
area $a_k$.  Contrary to the case of discs (or other constant shapes),
these three quantities are independent. So there is no obvious order
of the $A_k$.  Let $\{A_k\}_{k=1}^\infty$ be such that the total area
$\sum_{k=1}^\infty a_k$ is equal to the area of $A$. Then the set $E =
A - \cup_{k=1}^\infty A_k$ has zero area.  To calculate its box
dimension, we can use the Minkowski-Bouligand dimension
\cite{Mandel,Falc2}, which is equivalent to the box dimension
\cite{Falc2}.

The Minkowski-Bouligand dimension expresses the power-law behaviour of
the $r$-neigh\-bour\-hood of $E$ as $r \ra 0$.  To apply it to the
cut-out set $E$, one needs to distinguish voids that are included in
the $r$-neighbourhood of $E$ from voids that are not included. In any
one of the latter, the $r$-neighbourhood of $E$ is a band of width $r$
that leaves an empty part. We can try to estimate the areas of these
bands, in order to obtain the area of the $r$-neighbourhood of
$E$. Before doing so, let us recall Falconer's formula for the area of
the $r$-neighbourhood of $E$ in the case of disc-shaped voids.

\subsubsection{Area of the $r$-neighbourhood of $E$ 
with disc-shaped voids}

Falconer's formula \cite{Falc} for the area of the $r$-neighbourhood
of $E$ in the case of disc-shaped voids reads:
\begin{eqnarray}
V(r) = (p\,r + \pi r^2) +
\sum_{i=1}^k \pi \left(r_i^2 - (r_i-r)^2 \right) +
\sum_{i=k+1}^\infty \pi\, r_i^2    \label{V1} \\
= p\,r + 2\pi r \sum_{i=1}^k r_i + 
\pi \sum_{i=k+1}^\infty r_i^2 + \pi\, r^2 (1-k), \label{V2}
\end{eqnarray}
where $k$ is such that $r$ is between the disc radii of indices $k$
and $k+1$ ($r_{k+1} \leq r \leq r_{k}$).  The first term of Eq.\
(\ref{V1}) is the area of a band surrounding $A$, the second term is
the area of the annuli of width $r$ inside the discs $A_i$ for $1 \leq
i \leq k$, and the third term the area of the discs $A_i$ for $i \geq
k + 1$. In the second expression, Eq.\ (\ref{V2}), the second term,
which is the area of the annuli of width $r$ except for $-k\pi r^2$
(added to the last term), represents the sum of the perimeters of the
discs $A_i$ for $1 \leq i \leq k$ multiplied by $r$.

Falconer assumes that the disc radii and, therefore, areas fulfill
Zipf's law, with a concrete formulation, namely, $r_k \asymp k^{-\a}$,
in terms of the relation $\asymp$.  The relation $a_k \asymp b_k$
between two sequences $\{a_k\}$ and $\{b_k\}$ 
means that there are two constants $c_1, c_2 > 0$ such that $c_1
\leq a_k/b_k \leq c_2$ for all $k$ (this concept also applies to
functions, having the relation $f(x) \asymp g(x)$ analogous meaning).
Falconer further assumes that $1/2 < \a < 1$. The inequality $\a >
1/2$ ensures that the series of disc areas converges. The role of the
inequality $\a < 1$ is more subtle: it implies that the series of disc
perimeters diverges.  So, for $r_{k+1} \leq r \leq r_{k}$, we have
\begin{eqnarray}
V(r) \asymp r + r \sum_{i=1}^k i^{-\a} + 
\sum_{i=k+1}^\infty i^{-2\a} - r^2 k  \asymp 
r\,k^{1-\a} + k^{1-2\a} \asymp r^{2-1/\a},   %\label{V2}
\end{eqnarray}
and 
$$\dim_B E = 2 - \lim_{r \ra 0} \frac{\log V(r)}{\log r} =
\frac{1}{\a} > 1.$$ 
If $\a > 1$ the series of perimeters converges and
the second term in Eq.\ (\ref{V2}) behaves like $r$, dominating over the
terms that behave like $r^{2-1/\a}$. This implies that $\dim_B E = 1$,
instead of $\dim_B E = 1/\a < 1$. We encounter again that a cut-out
set must have dimension larger than $d-1$, since this is the
topological dimension of the boundaries of voids.

\subsubsection{Area of the $r$-neighbourhood of $E$ 
with triangular voids}

Before proceeding to analysing the case of general convex polygonal
voids, let us analyse the simpler case of triangles (which are
necessarily convex).

Let us consider voids that are not included in the $r$-neighbourhood
of $E$.  In any of these voids, the $r$-neighbourhood of $E$ forms a
band of width $r$ that leaves an empty part, which we can
calculate. To be precise, the $r$-neighbourhood of the boundary of a
triangle is a band of width $r$ so the region uncovered is another
triangle with the same angles and, therefore, similar.  The similarity
ratio can be determined employing some elementary geometry.  The
vertices of the similar triangles for various $r$ are placed along the
angle bisectors and so the triangles themselves are homothetical with
respect to the common centre of their inscribed circles. The area of
any of these triangles is given in terms of its perimeter $p$ and the
radius of its inscribed circle $\r$ by $a = p\r/2$. If $\r$ refers to
the original triangle and $\r'$ to another, the similarity ratio is
$\r'/\r$. So the area of the latter is $a' = a (\r'/\r)^2$ and $a - a'
= a [1-(\r'/\r)^2]$. Given that $r = \r - \r'$, we obtain for the area
of the band of width $r$
\begin{eqnarray}
a - a' = p\,r - \frac{p\,r^2}{2\r} = p\,r - \frac{p^2r^2}{4a}. 
\label{a-a}
\end{eqnarray}
In this formula, the factor $p^2/(4a)$ of $r^2$ only depends on the
angles and is a measure of shape.  The value of $a'$ decreases as $r$
increases and vanishes for the maximal $r = \r$.

The preceding results show us the appropriate version of formulas
(\ref{V1}) and (\ref{V2}) for the area of the $r$-neighbourhood of
$E$:
\begin{eqnarray}
V(r) = (p\,r + \pi r^2)  + 
\sum_{i=1}^k 
\left(p_i r - \frac{p_i^2 r^2}{4 a_i} \right) +
\sum_{i=k+1}^\infty a_i    \label{Vt1} \\
= p\,r + r \sum_{i=1}^k p_i + 
\sum_{i=k+1}^\infty a_i + r^2 \left(\pi - 
\sum_{i=1}^k \frac{p_i^2}{4 a_i} \right), \label{Vt2}
\end{eqnarray}
where $k$ is such that ${\r_{k+1}} \leq r \leq {\r_{k}}.$
To proceed like in the case of discs, we would like to have
$$\r_k \asymp k^{-\a}, \; p_k \asymp \r_k, \; a_k \asymp \r_k^2\,.$$
So we must have that $a_k \asymp p_k^2$, that is, that the perimeter
to area ratio $p^2/(4a)$ is bounded above and below.  According to the
basic {\em isoperimetric inequality}, this quantity has an absolute
lower bound of $\pi$, reached by a disc (for triangles the lower bound
is larger, namely, $3\sqrt{3}$, reached by the equilateral triangle)
\cite{geom}.  The upper bound must be explicitly imposed. Then, from
$\r_k \asymp k^{-\a}$ and $a_k \asymp p_k^2$ we deduce $p_k \asymp
k^{-\a}$. Therefore, a proof analogous to Falconer's proof for disc
voids leads to the same result, namely, $\dim_B E = 1/\a$ ($1/2 < \a
<1$).

An upper bound to the perimeter to area ratio $p^2/(4a)$ is an
intuitively reasonable requirement. This ratio only depends on the
angles of the triangle and an upper bound to it is equivalent to a
lower bound to them. In other words, we are excluding ``spiky"
triangles, which are nearly one-dimensional and can be packed in a
small area without reducing their diameter.

\subsubsection{Area of the $r$-neighbourhood of $E$ 
with convex polygonal voids}

In this case, in a void convex polygon not included in the
$r$-neighbourhood of $E$, the $r$-neighbourhood also forms a band of
width $r$ that leaves an empty part, but its area is harder to
calculate.  Therefore, instead of attempting to derive an equation
analogous to Eq.\ (\ref{Vt2}), we look for independent bounds to
$V(r)$. A lower bound is certainly provided by the sum of the areas of
polygons fully included in the $r$-neighbourhood.  An upper bound to
$V(r)$ requires us to estimate the area of bands of with $r$ inside
larger polygons and will be given by an expression simpler than the
right-hand side of Eq.\ (\ref{Vt2}).

To find a precise expression of the lower bound to $V(r)$, we need a
criterium to determine when a void polygon is fully included in the
$r$-neighbourhood of $E$.  A simple criterium is given by the diameter
$\d$ of the polygons: if $r \geq \d$, then the void polygons of
diameter equal or smaller than $\d$ are certainly covered. So, if we
order the void polygons by their diameter and $k$ is such that
$\d_{k+1} \leq r \leq \d_k$,
\begin{eqnarray}
\sum_{i=k+1}^\infty a_i \leq V(r). 
\end{eqnarray}

To find a precise expression of the upper bound to $V(r)$, we need a
criterium to determine when a void polygon is {\em not} fully included
in the $r$-neighbourhood of $E$.  Take a particular polygon $A_i$,
with area $a_i$ and perimeter $p_i$. Given that the area of a band of
width $r$ inside $A_i$ and around its boundary is smaller than
$p_i\,r$, a sufficient condition is that $r \leq a/p$.  Therefore, an
upper bound to $V(r)$ is
\begin{eqnarray}
V(r) \leq pr + r \sum_{i=1}^k p_i + 
\sum_{i=k+1}^\infty a_i + \pi r^2\,, 
\label{Vp}
\end{eqnarray}
where the void polygons are ordered by their value of $a/p$ and $k$ is
such that
$$\frac{a_{k+1}}{p_{k+1}} \leq r \leq \frac{a_{k}}{p_{k}}\,.$$ We
again need the relation $a_k \asymp p_k^2$ to relate the second and
third terms of inequality (\ref{Vp}).  This condition implies a lower
bound to the angles of polygons, like for triangles.  However, the
condition is now stronger: it is possible to have ``spiky" convex
polygons (with large diameter to area ratio) with non-small angles; a
simple example is the rectangle, with no upper bound to the ratio
$p^2/(4a)$.

Then, let us impose the conditions 
$$p_k \asymp k^{-\a},\; a_k \asymp k^{-2\a}\,,$$ like we did for
triangles.  We have, for $a_{k+1}/p_{k+1} \leq r \leq a_{k}/p_{k},$
\begin{eqnarray}
V(r) \leq  c \left(r \sum_{i=1}^k i^{-\a} + 
\sum_{i=k+1}^\infty i^{-2\a} \right) + \pi r^2
\leq  c' (r\,k^{1-\a} + k^{1-2\a}) + \pi r^2 
\leq c'' r^{2-1/\a}, %\label{Vp}
\end{eqnarray}
for some positive numbers $c,\,c',\,c'',$ and $r < 1$.  On the other
hand, we can apply a similar procedure to the lower bound. To do so,
we first relate the diameter of a convex polygon with its perimeter.
We note that there are both lower and upper bounds to their ratio: $2
\d < p$ and $p \leq \pi\d$ (which actually hold for any {\em convex}
figure) \cite{geom}. So $\d_k \asymp p_k$.  Therefore, for $\d_{k+1}
\leq r \leq \d_k$,
$$c''' r^{2-1/\a} \leq V(r), $$ for some positive number $c'''$.  Both
bounds are equivalent to
$$ V(r) \asymp r^{2-1/\a},$$
implying that $\dim_B E = 1/\a$ ($1/2 < \a <1$).

In conclusion, the conditions 
\begin{equation}
a_k \asymp k^{-2\a}, \; a_k \asymp p_k^2 
\label{Zipf2}
\end{equation}
seem to be as suitable for convex polygonal voids as for triangular
voids or discs. In fact, an approximation argument would show that
these conditions are suitable for general convex voids.  The rationale
for this proof is that the quotient $p^2/a$ is the measure of shape
for convex figures and its being bounded ensures that they do not
degenerate in one-dimensional figures (segments), so that perimeter
and area have the natural scaling behaviour.

\subsection{Generalization to three-dimensional cut-out sets}

We proceed to the generalization to three-dimensional cut-out sets
with convex polyhedral voids.  We need to generalize the geometrical
properties that we have used from convex polygons to convex polyhedra.

Let $A$ be a compact convex region that is to become a cut-out set.
Falconer's formula (\ref{V1}) for the area of the $r$-neighbourhood of
$E$ can be generalized to three dimensions (ball-shaped voids):
\begin{eqnarray}
V(r) = (ar + Hr^2 + \frac{4}{3}\pi r^3)  + 
\sum_{i=1}^k \frac{4}{3}\,\pi \left(r_i^3 - (r_i-r)^3 \right) +
\sum_{i=k+1}^\infty \frac{4}{3}\,\pi r_i^3    \label{V1-3} \\
= (ar + Hr^2) + 4\pi r \sum_{i=1}^k r_i^2 - 
4\pi r^2 \sum_{i=1}^k r_i + 
\frac{4}{3}\,\pi \sum_{i=k+1}^\infty r_i^3 + 
\frac{4}{3}\pi r^3 (1+k), 
\label{V2-3}
\end{eqnarray}
where $k$ is such that $r$ is between the ball radii of indices $k$
and $k+1$ ($r_{k+1} \leq r \leq r_{k}$).  The first term of Eq.\
(\ref{V1-3}) is the volume of the layer surrounding $A$, given by
Steiner's formula, where $H$ is $A$'s {\em linear measure} (mean
curvature) \cite{geom}.  The second term is the volume of the shells
of width $r$ inside the balls $A_i$, for $1 \leq i \leq k$, and the
third term the area of the balls $A_i$ for $i \geq k + 1$. In the
second expression, Eq.\ (\ref{V2-3}), the second term represents the
sum of the areas of the balls $A_i$ for $1 \leq i \leq k$ multiplied
by $r$, while the third term represents the sum of their linear
measures multiplied by $r^2$.

We assume that $r_k \asymp k^{-\a}, \; 1/3 < \a < 1/2$, ensuring that
the series of ball areas diverges while the series of their volumes
converges.  So, for $r_{k+1} \leq r \leq r_{k}$, we have
\begin{eqnarray}
V(r) \asymp r + r^2 +r \sum_{i=1}^k i^{-2\a} + 
\sum_{i=k+1}^\infty i^{-3\a} + r^3 k  \asymp 
r\,k^{1-2\a} + k^{1-3\a} + r^3 k \asymp r^{3-1/\a},   %\label{V2}
\end{eqnarray}
and 
$$\dim_B E = 3 - \lim_{r \ra 0} \frac{\log V(r)}{\log r} =
\frac{1}{\a} > 2.$$ Again, a cut-out set must have dimension larger
than $d-1$ (the topological dimension of the boundaries of voids).

For tetrahedral voids, the $r$-neighbourhood of the boundary of a
tetrahedron is a layer of thikness $r$ and its volume is (in analogy
with Eq.\ (\ref{a-a}))
\begin{eqnarray}
v - v' = ar - \frac{a^2r^2}{3v} + \frac{a^3r^3}{27 v^2} \,.
\label{a-a-3}
\end{eqnarray}
Here $a$ and $v$ are the area surface and volume of the tetrahedral
void, respectively, and $v'$ is the volume of the smaller homothetical
tetrahedron.  The factor $a^3/(27v^2)$ only depends on the angles of
the tetrahedron and is a measure of its shape.  The maximal $r$, such
that $v'$ vanishes, is $r = \r = 3v/a$, that is, the radius of the
inscribed sphere.  The appropriate relations for tetrahedral voids are
$$\r_k \asymp k^{-\a}, \; a_k \asymp \r_k^2, \; 
v_k \asymp \r_k^3\,.$$
So the relation between surface area and volume to be extended to
general convex polyhedra is $v_k \asymp a_k^{3/2}$. The lower bound to
$a^{3/2}/v$ is again universal, according to a three-dimensional
isoperimetric inequality, and corresponds to a ball (for tetrahedra,
to the regular tetrahedron). The upper bound forbids again small
angles that give rise to flattened or spiky tetrahedra. Note that a
tetrahedron can degenerate either into a two-dimensional or a
one-dimensional figure (triangle or segment, respectively), but the
former is more generic.

In the case of convex polyhedral voids, in analogy with two
dimensions, a lower bound to $V(r)$ is provided by the sum of the
volumes of polyhedra fully included in the $r$-neighbourhood and an
upper bound is given by the three-dimensional version of Eq.\
(\ref{Vp}).  For the lower bound to $V(r)$, to determine when a void
polyhedron is fully included in the $r$-neighbourhood of $E$, we use
again the criterium given by the diameter $\d$ of the polyhedra,
namely, $r \geq \d$. So, if we order the void polyhedra by their
diameter and $k$ is such that $\d_{k+1} \leq r \leq \d_k$,
\begin{eqnarray}
\sum_{i=k+1}^\infty v_i \leq V(r). 
%\label{Vp}
\end{eqnarray}

For the upper bound to $V(r)$ we need a condition that ensures that a
given polyhedron is not included in the $r$-neighbourhood of $E$.  The
volume of the layer of thikness $r$ inside a polyhedron $A_i$ is
smaller than $a_i\,r$.  We have (in analogy with inequality (\ref{Vp})
for polygons):
\begin{eqnarray}
V(r) \leq ar + H r^2 + r \sum_{i=1}^k a_i + 
\sum_{i=k+1}^\infty v_i + 
\frac{4}{3}\pi r^3 
%\label{Vp}
\end{eqnarray}
where the void polyhedra are ordered by their value of $v/a$ and $k$
is such that
$$\frac{v_{k+1}}{a_{k+1}} \leq r \leq \frac{v_{k}}{a_{k}}\,.$$
Assuming that 
\begin{equation}
v_k \asymp k^{-3\a}, \; v_k \asymp a_k^{3/2}\,,
\label{Zipf3}
\end{equation}
it follows that
$$V(r) \leq c \,r^{3-1/\a},$$
for some positive number $c$, and $r < 1$.

For the lower bound to $V(r)$, we may relate the diameter of a convex
polyhedron with its linear measure $H$.  Like in the two-dimensional
case, there are both lower and upper bounds to their ratio: $\d <
H/\pi$ and $H \leq 2\pi\d$, which hold for any {\em convex} body
\cite{geom}.  $H$ is {\em independent} of $v$ and $a$, 
but the relations (\ref{Zipf3}) imply $H_k \asymp k^{-\a}$
nonetheless.  This follows from the two fundamental three-dimensional
isoperimetric inequalities: $a^2 \geq 3vH,\; H^2 \geq 4\pi a$
\cite{geom}.  Therefore, for $\d_{k+1} \leq r \leq \d_k$,
$$c' r^{3-1/\a} \leq V(r), $$
for some positive number $c'$.
Both upper and lower-bound inequalities are equivalent to
$$ V(r) \asymp r^{3-1/\a},$$
so $\dim_B E = 1/\a$ ($1/3 < \a <1/2$).

In conclusion, the conditions (\ref{Zipf3}) are suitable for convex
polyhedral voids and, furthermore, an approximation argument would
show that they are suitable for general convex voids.  Note that our
proof relies on the sufficiency of the upper bound to the quotient
$a^3/v^2$ (as a measure of shape) for preventing degeneracy into lower
dimension.  This holds regardless of the actual existence of two
independent measures of shape of three-dimensional convex bodies.

\section{Large scale distribution of matter and cosmic voids}
\label{cosmo}

The large scale distribution of matter in cosmology is produced by the
gravitational instability of primordial small fluctuations in a
homogenous Friedman-Robertson-Walker universe \cite{Pee,Padma}. The
dynamics of structure formation is very nonlinear and, therefore,
difficult to study with analytic methods. However, this dynamics is
scale invariant, at least within some range of scales, due to the
scale invariance of gravity.  In consequence, it is natural that a
scale invariant distribution of matter develops, with fractal
geometry.  Indeed, the fractal geometry of the distribution of
galaxies has been studied for years \cite{Piet-Marti}, and the study
of galaxy clustering actually stimulated the development of fractal
geometry \cite{Mandel}.

Mandelbrot considered in his book \cite{Mandel} the presence of voids
in the distribution of galaxies but, according to the observational
situation at the time of writing it, favored small voids and,
actually, introduced the concept of {\em lacunarity} to account for
this feature. The observation of large voids in the distribution of
galaxies is more recent.  Surprisingly, the cosmological literature on
voids \cite{cosmo} hardly treats their fractal properties. Trying to
fill this gap, we began a program to adapt the algorithmic studies of
cosmic voids to general fractals and, viceversa, to discern fractal
features in cosmic voids \cite{Gaite,Gaite2}.  In particular, we
proposed to employ standard rank-ordering techniques and test the
already compiled catalogues of cosmic voids for Zipf's law.  In Ref.\
\cite{Gaite2}, we devised a void-finding algorithm based on discrete
geometry methods, namely, Delaunay and Voronoi tessellations, that
produces a sequence of voids of polyhedral shape (in $d=3$).

The concepts of discrete stochastic geometry had been introduced in
cosmology before by Rien van de Weygaert and collaborators, with
different purposes (a comprehensive reference is \cite{Rien}). One of
these purposes was actually the construction of a model of large-scale
structure formation based on Voronoi tessellations. In this model,
Voronoi cells represent void regions while the matter is concentrated
in their walls. The cell centers represent void germs (contrary to
their role in our void-finding algorithm, in which they correspond to
matter particles).

It is pertinent here to mention that there is a successful model of
large-scale structure formation that produces walls as first
structures, namely, the adhesion model \cite{Padma} (the walls are
called ``pancakes" in this context).  The full structure produced by
this model is a self-similar pattern of interlocking walls that has
been dubbed the ``cosmic web" (or the ``cosmic foam") \cite{Rien}.  As
matter concentrate in the walls, there appear depleted regions, that
is, voids. On account of the self-similarity, the voids form a {\em
hierarchy}, akin to the void distribution given by Zipf's law. Most
studies of the adhesion model focus on the distribution of matter and
its evolution, but it has been argued that it makes more sense to
focus on the evolution of underdense regions.  The rationale for this
viewpoint is the ``bubble theorem" \cite{Rien}: the evolution of an
underdense region is such that its initial slight asphericity
decreases, contrary to the evolution of an overdense
region. Therefore, the evolution of underdense regions is essentially
decribed by the expansion of ellipsoidal voids, which become more
spherical until they collapse with other voids, forming walls.

\subsection{The Voronoi foam model}

This model was proposed by Icke and Van de Weygaert \cite{Rien} and
follows the idea of expansion of ellipsoidal voids.  It is based on a
set of points that are initially the peaks of the gravitational
potential, where the matter is underdense. They become the ``expansion
centres" of matter flowing outwards with uniform velocity. Voids form
in this manner. Furthermore, when the flow from one void encounters
the flow from an adjacent one, a wall forms half-way between their
centres. The resulting distribution is a set of Voronoi cells, that
is, a ``Voronoi foam". These Voronoi cells are convex
polyhedra. Moreover, if we assume that they form a self-similar
pattern, then these patterns are particular cases of cut-out sets with
convex polyhedral voids.

Unfortunately, the Voronoi foam model, such as has been formulated, is
not sufficient to deduce the similarity properties of the cell
pattern. If we understand the ``expansion centres" as the set of
relative maxima of the gravitational potential, we have to consider
that this set is {\em dense} for an initial random Gaussian
distribution.  So hardly any void expansion seems possible, unless one
selects a subset of ``expansion centres", according to some principle,
which will determine the final cell pattern.  Of course, an obvious
condition for a self-similar cell pattern is that the cell sizes
fulfill Zipf's law, namely, the rank-ordering of their volumes
satisfies $v_n \asymp n^{-e}$ for some $e$.  To ensure that the
cut-out set defined in this manner be a fractal, in the sense of
having a box dimension strictly between two and three, we have to
demand $1 < e < 3/2$ and the condition that forbids the appearance of
{\em degenerate} (quasi-planar) shapes.  Provided that the
scale-invariant dynamics implies Zipf's law for the void cells, with
$1 < e < 3/2$, the ``bubble theorem" must imply that they are
non-degenerate.

\section{Discussion}

We have proved the relation between the box dimension of a cut-out set
$E$ and the exponent of the corresponding Zipf law, for convex voids,
in particular, for convex polygonal voids in $d=2$ and convex
polyhedral voids in $d=3$. The particular forms of Zipf's law for
voids in $d=2$ and $d=3$ are Eqs.\ (\ref{Zipf2}) and (\ref{Zipf3}),
respectively.  So we have extended Falconer's results in $d=1$ and
$d=2$ (further extension to convex voids in any dimension $d$ seems
straightforward).  If the Zipf law exponent is $e$ ($= d\,\a$), the
relation is $\dim_B E = d/e$.  Sufficient conditions for this relation
to hold is that $1 < e < d/(d-1)$ and, in addition, the exclusion of
{\em degenerate} void shapes.  We expect that the case with more
physical applications is $d=3$.  In particular, we have explored the
application of our result to the large scale distribution of matter in
cosmology.

It is useful to make a few remarks on the box-dimension formula.  We
have emphasized that cut-out sets must have topological codimension
one, that is, topological dimension $d-1$.  This is the reason why $e
< d/(d-1)$, which implies $\dim_B E = d/e > d-1$, according to the
known order of box and topological dimensions: if $e > d/(d-1)$, the
$d-1$-measure of the boundaries of voids converges and $\dim_B E =
d-1$.  On the other hand, the order of box, Hausdorff-Besicovitch and
topological dimensions is $\dim_T E \leq
\dim_H E \leq \dim_B E$, so $e > 1$ implies that $\dim_H E <
d$. However, we cannot determine if the inequality $\dim_T E \leq
\dim_H E$ is strict, so we cannot tell if $E$ is in fact a fractal
(according to the usual definition). As a one-dimensional example,
consider the ``convergent sequence sets'' $E^{(p)} = \{0,1,2^{-p},
3^{-p},4^{-p}, \ldots\},\; p>0,$ with $\dim_B E^{(p)} = 1/(p+1)$
\cite{Falc}: they have all but one of their points isolated and
$\dim_H E^{(p)} = 0$. Let us recall that the box
dimension only depends on the size of the voids and not on their
arrangement, but the Hausdorff-Besicovitch dimension can be altered by
a rearrangement of the voids.

The mathematical requirement of having topological codimension one can
be difficult to test in point sets obtained from physical
observations, which are necessarily finite. In fact, the definition of
void itself becomes uncertain in a finite set of points. This is why
one must resort to void-finding algorithms. These algorithms may have
free parameters, giving rise to different sets of voids according to
their value. We noted in Ref.\ \cite{Gaite2} the possibility of
``percolation of voids'', which must be avoided, by selecting
parameter values that produce small convex voids.  In this way, we
expect, when the number of sampling points of a cut-out fractal set
grows, that the set of voids found approaches the real set of
voids. However, questions of convergence are difficult to treat in a
rigorous way.  If a fractal set $E$ has $\dim_B E > d-1,$ but has
topological codimension larger than one, the cut-out set defined by
the set of voids found by some algorithm is a fractal set $E' \supset
E$ with the same box dimension $\dim_B E' = \dim_B E$, {\em and} with
topological codimension one. Naturally, the boundary of the voids
includes $E' - E$, with $\dim_H (E'-E) \leq \dim_B (E'-E) \leq d-1$,
so it is negligible with respect to $E$.  We have mentioned in section
\ref{Cantor} an example belonging to Cantor-like fractals.

Regarding the application of our results to the large scale structure
of matter, we can rely on both observational and theoretical results
to support the existence of voids and scale invariance. The
observations refer to the galaxy positions, rather than to the full
dark matter distribution. For the moment, in spite of the presence of
voids in the galaxy distribution, it is a moot point whether or not
the galaxies are distributed along walls (even though the expression
``wall galaxy" is in use, especially in the cosmological literature
about voids). It is even more uncertain whether or not the voids
satisfy Zipf's law \cite{Gaite}. The application of various
void-finders to galaxy catalogues yields results that are not
necessarily consistent. In contrast, the available theories of
non-linear gravitational clustering support the existence and scaling
of voids; in particular, the adhesion model leads to the formation of
a self-similar ``cosmic foam'', as commented in Sect.\ \ref{cosmo}. In
this context, the relation obtained here between the Zipf law for
voids and box and Hausdorff-Besicovitch dimensions will surely be
helpful.

\subsection*{Acknowledgments}
I thank M.\ Santander and D.\ Alarcos for a conversation on integral
geometry.  My work is supported by the ``Ram\'on y Cajal'' program and
by grant BFM2002-01014 of the Ministerio de Edu\-caci\'on y Ciencia.

%\newpage

\end{document}